\documentclass[aps,pra,twocolumn,showpacs,floatfix]{revtex4}
\usepackage{graphicx}
\begin{document}
\title{Ultracold collisions in magnetic fields: \\
reducing inelastic cross sections near Feshbach resonances}
\author{R. Adam Rowlands}
\affiliation{Department of Chemistry, University of Durham, South
Road, Durham, DH1~3LE, England}
\author{Maykel Leonardo
Gonz{\'a}lez-Mart{\'\i}nez} \affiliation{Departamento de
F{\'\i}sica General, InSTEC, Habana 6163, Cuba}
\author{Jeremy M. Hutson}
\affiliation{Department of Chemistry, University of Durham, South
Road, Durham, DH1~3LE, England}

\date{\today}

\begin{abstract}
We have carried out bound-state and low-energy quantum scattering
calculations on He + NH ($^3\Sigma^-$) in magnetic fields, with
the NH molecule in its $n=1$ rotationally excited states. We have
explored the pattern of levels as a function of magnetic field and
identified the nearly good quantum numbers in different regimes.
We have used the bound-state calculations to locate low-energy
Feshbach resonances. When the magnetic field is used to tune
across such a resonance, the real and imaginary part of the
scattering length show asymmetric oscillations and peaks with
amplitude between 1 and 3 \AA. The scattering length does {\it
not} pass through a pole at resonance. The resonant behavior is
characterized by a complex resonant scattering length $a_{\rm
res}$. The corresponding inelastic cross sections show troughs as
well as peaks near resonance. This may be important for efforts to
achieve evaporative and sympathetic cooling for molecules, because
it offers the hope that inelastic trap losses can be reduced by
tuning close to a Feshbach resonance.
\end{abstract}

\pacs{34.50.-s,34.10.+x,03.65.Nk,82.20.Xr,34.30.+h}

\maketitle


\section{Introduction}

Collisions between ultracold atoms can be controlled by tuning the
atomic interactions using applied magnetic fields
\cite{Hutson:IRPC:2006, Koehler:RMP:2006}. Such techniques have
been used to cause controlled implosion of Bose-Einstein
condensates \cite{Roberts:2001} and to produce dimers of both
bosonic \cite{Donley:2002, Herbig:2003, Xu:2003,
Durr:mol87Rb:2004} and fermionic \cite{Regal:40K2:2003,
Strecker:2003, Cubizolles:2003, Jochim:Li2pure:2003} alkali metal
atoms. Long-lived molecular Bose-Einstein condensates of fermion
dimers have been produced \cite{Jochim:Li2BEC:2003,
Zwierlein:2003, Greiner:2003}, and the first signatures of
ultracold triatomic \cite{Kraemer:2006} and tetraatomic
\cite{Chin:2005} molecules have been observed. It is proving
possible to move molecules selectively between quantum states by
either fast or slow sweeps of magnetic fields across avoided
crossings between bound states \cite{Mark:stuck:2007,
Mark:spect:2007}.

The overall strength of the interaction between a pair of atoms is
characterized by the scattering length $a$ \cite{Pethick:2002}. An
important feature of elastic scattering in ultracold atomic gases
is that $a$ passes through a pole as the magnetic field $B$ is
swept across a Feshbach resonance at constant kinetic energy
\cite{Moerdijk:1995},
\begin{equation}
a(B) = a_{\rm bg} \left[ 1 - \frac{\Delta_B}{B - B_{\rm res}}
\right], \label{eqares}
\end{equation}
where $a_{\rm bg}$ is a near-constant background scattering length
and $B_{\rm res}$ and $\Delta_B$ are the position and width of the
resonance. The scattering length can thus be tuned to any desired
value, positive or negative. Positive values correspond to
interactions that are overall repulsive and negative values to
interactions that are overall attractive. The elastic cross
section is given by
\begin{equation}
\sigma_{\rm el} = \frac{4\pi a^2}{1+k^2a^2}, \label{eqsigel1}
\end{equation}
where the kinetic energy is $E_{\rm kin}=\hbar^2 k^2/(2\mu)$ and
$\mu$ is the reduced mass for the collision. The elastic cross
section thus passes through a peak of height $4\pi/k^2$ at
resonance.

It has recently become possible to cool molecules directly from
high temperature to the millikelvin regime, using methods such as
buffer-gas cooling \cite{Weinstein:CaH:1998, Egorov:2004,
Campbell:2007} and Stark deceleration \cite{Bethlem:IRPC:2003,
Bethlem:2006}. Polar molecules such as ND$_3$ and OH have been
successfully trapped at temperatures around 10 mK, and there are a
variety of proposals for ways to cool them further, including
evaporative cooling, sympathetic cooling and cavity-assisted
cooling \cite{Domokos:2002, Chan:2003, Morigi:2007}. Very
recently, NH has been trapped at temperatures around 0.7 K by
buffer-gas cooling in cryogenic helium.

In previous work \cite{Gonzalez-Martinez:2007}, we have explored
the possibility of controlling {\it molecular} interactions in the
same way as atomic interactions. We have generalised the BOUND
\cite{Hutson:bound:1993} and MOLSCAT \cite{molscat:v14} packages
to carry out bound-state calculations and quantum scattering
calculations in applied magnetic fields for systems made up of a
$^{2S+1}\Sigma$ molecule and a structureless atom.

In our initial calculations on He + NH ($^3\Sigma$)
\cite{Gonzalez-Martinez:2007}, we used BOUND to locate magnetic
fields at which bound states cross open-channel thresholds. We
then used MOLSCAT to characterize the resulting low-energy
Feshbach resonances as a function of magnetic field. For a
resonance at which a bound state crossed the {\it lowest}
open-channel threshold, we observed a pole in the scattering
length that followed Eq.\ \ref{eqares}. However, for a resonance
in which a state crossed a {\it higher} threshold, we observed
quite different behavior. The scattering length showed only a weak
oscillation instead of a pole. The suppression of the pole was
attributed to inelastic effects. The calculations were for NH in
its lowest rotational state ($n=0$), for which inelastic coupling
is very weak, and the resonances were very narrow, but even so the
amplitude of the oscillation in $a(B)$ was only about 9 \AA.

A full derivation of the resonant behavior of the scattering
length in the presence of inelastic effects has been given
previously \cite{Hutson:res:2007}, so only a brief version will be
given here to explain the basic physics and define notation. In
the presence of inelastic collisions, the scattering matrix $S$
that describes the collision in quantum-mechanical terms has
elements $S_{ii'}$. The diagonal S-matrix element in the incoming
channel $0$ has magnitude $|S_{00}|\le 1$ and may be written in
terms of a complex phase shift $\delta_0$ with a positive
imaginary part \cite{Mott:p380:1965},
\begin{equation}
S_{00}(k_0)=e^{2{\rm i}\delta_0(k_0)}, \label{eqsd}
\end{equation}
where $k_0$ is the wave vector in the incoming channel. This can
be expressed in terms of a complex energy-dependent scattering
length, $a(k_0)=\alpha(k_0)-{\rm i}\beta(k_0)$ \cite{Bohn:1997,
Balakrishnan:scat-len:1997}, defined as
\begin{equation}
a(k_0)= \frac{-\tan\delta_0(k_0)}{k_0} = \frac{1}{{\rm i}k_0}
\left(\frac{1-S_{00}(k_0)}{1+S_{00}(k_0)}\right). \label{eqacomp}
\end{equation}
$a(k_0)$ becomes constant at limitingly low kinetic energy. The
elastic and total inelastic cross sections are exactly
\cite{Cvitas:li3:2007}
\begin{equation}
\sigma_{\rm el}(k_0) =
\frac{4\pi|a|^2}{1+k_0^2|a|^2+2k_0\beta} \label{eqsigela}
\end{equation}
and
\begin{equation}
\sigma_{\rm inel}^{\rm tot}(k_0) =
\frac{4\pi\beta}{k_0(1+k_0^2|a|^2+2k_0\beta)}. \label{eqsiginela}
\end{equation}

A scattering resonance is most simply characterized in terms of
the S-matrix eigenphase sum $\Sigma$, which is the sum of phase
shifts obtained from the eigenvalues of the S matrix
\cite{Hazi:1979, Ashton:1983}. The eigenphase sum is a real
quantity, and across a resonance follows the Breit-Wigner form,
\begin{equation}
\Sigma(B) = \Sigma_{\rm bg} + \tan^{-1}
\left[\frac{\Gamma_B}{2(B_{\rm res}-B)}\right], \label{eqbwb}
\end{equation}
where $\Sigma_{\rm bg}$ is a slowly varying background term,
$B_{\rm res}$ is the resonance position and $\Gamma_B$ is a
resonance width (not the same as $\Delta_B$ in Eq.\ \ref{eqares}).
The $B$ subscripts indicate that we are considering the resonance
as a function of magnetic field rather than energy. The individual
S-matrix elements describe circles in the complex plane
\cite{Brenig:1959, Taylor:p411:1972, Gazdy:1987, Hutson:res:2007},
\begin{equation}
S_{ii'}(B) = S_{{\rm bg,}ii'} - \frac{{\rm i} g_{Bi} g_{Bi'}}{B -
B_{\rm res} + {\rm i}\Gamma_B/2}, \label{eqsii}
\end{equation}
where $g_{Bi}$ is complex. The radius of the circle in $S_{ii}$ is
$|g_{Bi}^2/\Gamma_B|$. The {\it partial width} $\Gamma_{Bi}$ for
channel $i$ is usually defined as a real quantity, but here we
also need a corresponding phase $\phi_i$ to describe the {\it
orientation} of the circle in the complex plane,
$g_{Bi}^2=\Gamma_{Bi} e^{2{\rm i}{\phi_i}}$. The width $\Gamma_B$
and partial widths $\Gamma_{Bi}$ are signed quantities, positive
if the resonant state tunes downwards across the threshold as a
function of $B$ and negative if it tunes upwards. For a narrow
resonance, the total width is just the sum of the partial widths,
\begin{eqnarray} \Gamma_B = \sum_i \Gamma_{Bi}.
\end{eqnarray}

\begin{figure}
\includegraphics[width=0.95\linewidth]{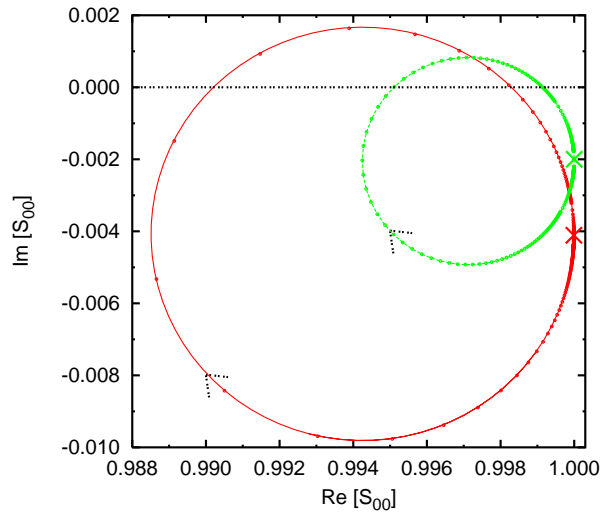}
\caption{(Color online). The small circles in the elastic S-matrix
elements in the presence of inelastic scattering for He + NH
($n=0$) scattering at $E_{\rm kin}=10^{-6}$~K (green, smaller
circle) and $4 \times 10^{-6}$~K (red, larger circle) Reproduced
with permission from ref.\ \onlinecite{Gonzalez-Martinez:2007}.}
\label{figsm-n0}
\end{figure}

The partial widths for {\it elastic} channels (degenerate with the
incoming channel) are proportional to $k_0$ at low energy. We may
define a reduced partial width $\gamma_{B0}$ for the incoming
channel by
\begin{eqnarray}
\Gamma_{B0}(k_0) &=& 2k_0\gamma_{B0}, \label{eqgamlin}
\end{eqnarray}
and the reduced width is independent of $k_0$ at low energy
(typically below $E_{\rm kin} = 1$ mK). By contrast, the partial
widths for {\it inelastic} channels depend on open-channel
wavefunctions with large wave vectors $k_i$ and are effectively
independent of $k_0$ in the ultracold regime. If the inelastic
partial widths $\Gamma_{Bi}$ are non-zero, they eventually
dominate $\Gamma_{B0}$ as $k_0$ decreases. The radius of the
circle (\ref{eqsii}) described by $S_{00}$ thus drops linearly to
zero as $k_0$ decreases, as shown in Figure \ref{figsm-n0}. This
is {\it qualitatively different} from the behavior in the absence
of inelastic channels, where the circle has radius 1 even at
limitingly low energy.

\begin{figure}
\begin{center}
\includegraphics[width=0.95\linewidth]{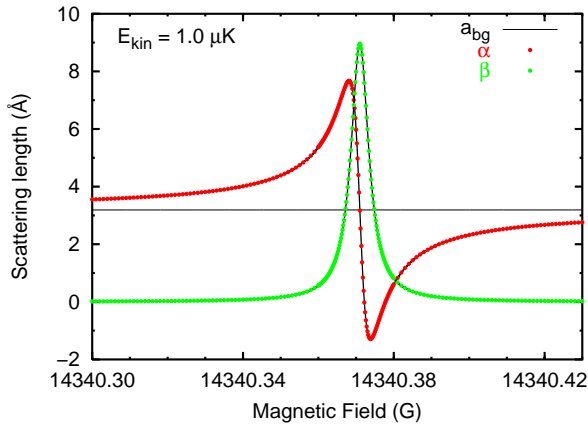}
\end{center}
\caption{(Color online). Real and imaginary parts of the
scattering length across a Feshbach resonance in He + NH ($n=0$),
showing a small symmetrical oscillation in the real part (red) and
a peak in the imaginary part (green). Reproduced with permission
from ref.\ \onlinecite{Hutson:res:2007}.} \label{fig-HeNH-a-n0}
\end{figure}

\begin{figure}
\includegraphics[width=0.95\linewidth]{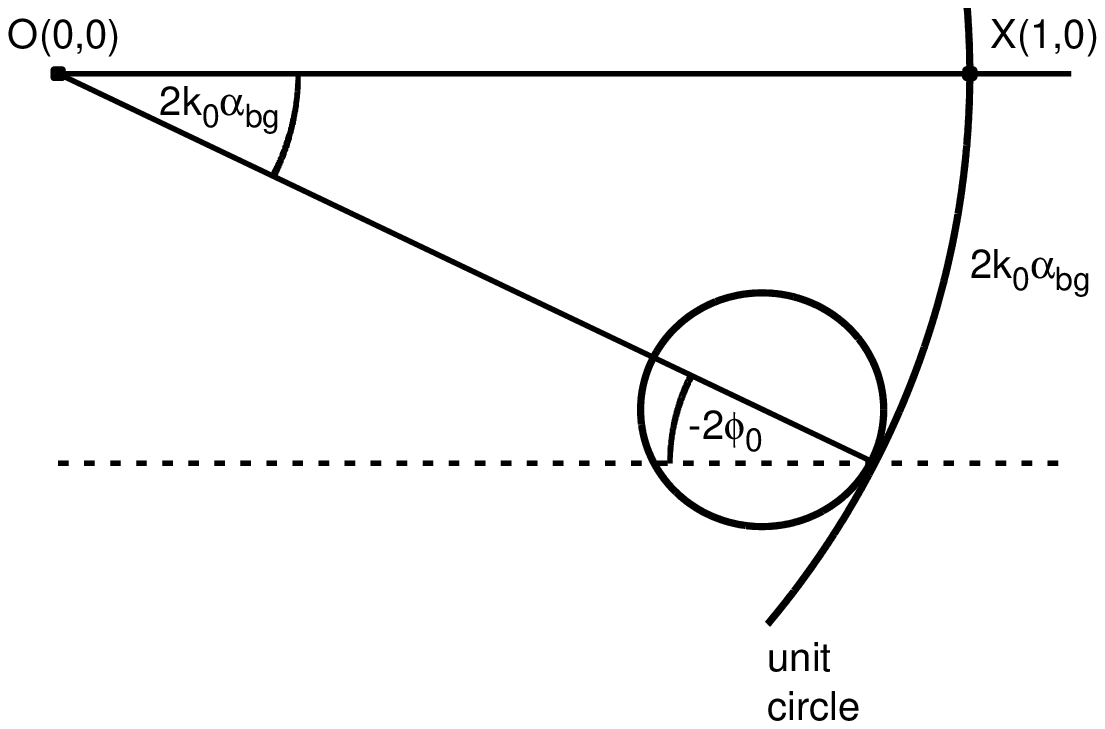}
\includegraphics[width=0.95\linewidth]{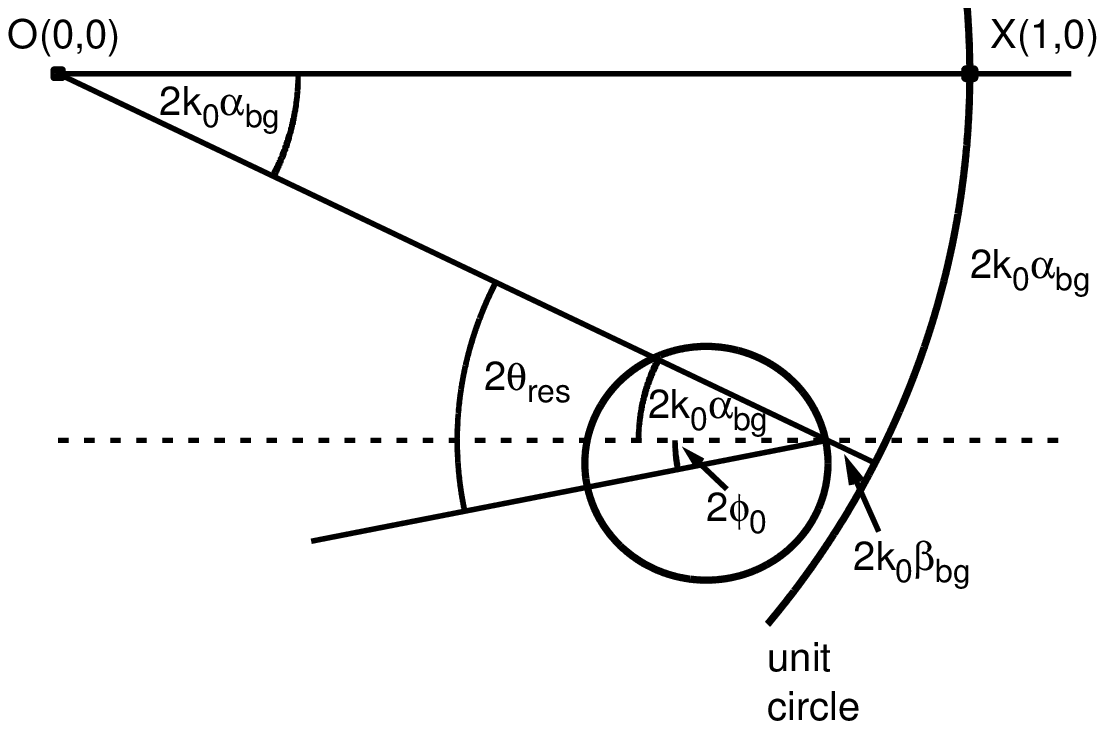}
\caption{The distinction between an S-matrix circle in the complex
plane that points directly towards the origin O (top) and one that
does not (bottom), showing the relationship between $\theta_{\rm
res}$ and $\phi_0$.} \label{fig-s-circles}
\end{figure}

The radius of the circle in $S_{00}$ is $\Gamma_{B0}/\Gamma_B$.
For small $k_0$, where Eq.\ \ref{eqgamlin} applies, this is
approximately $2k_0 \gamma_{B0}/\Gamma_B^{\rm inel}$. The formula
followed by the complex scattering length is
\begin{equation}
a(B) = a_{\rm bg} + \frac{a_{\rm res}}{2(B-B_{\rm
res})/\Gamma_B^{\rm inel}+{\rm i}}, \label{eqaares}
\end{equation}
where $\Gamma_B^{\rm inel}$ is the energy-independent part of
$\Gamma_B$ (omitting $\Gamma_{B0}$) and $a_{\rm res}$ is a {\it
resonant scattering length} that characterizes the strength of the
resonance,
\begin{equation}
a_{\rm res}=\frac{2\gamma_{B0}}{\Gamma_B^{\rm inel}} \, e^{2{\rm
i}(\phi_0+k_0\alpha_{\rm bg})}. \label{eqaresdef}
\end{equation}
Both $a_{\rm res}$ and the background term $a_{\rm bg}$ can in
general be complex and are independent of $k_0$ at low energy.
However, in the special case where the background scattering is
purely elastic ($a_{\rm bg}$ is real), unitarity requires that the
circle in $S_{00}$ must loop towards the origin as shown in the
upper panel of Fig.\ \ref{fig-s-circles}. This requires that
$a_{\rm res}$ is also real. Across the width of the resonance, the
real part $\alpha(B)$ of the scattering length $a(B)$ then
oscillates about $a_{\rm bg}$ by $\pm a_{\rm res}/2$ and the
imaginary part peaks at $\beta(B)=a_{\rm res}$. This was the
behavior seen for $a(B)$ in ref.\
\onlinecite{Gonzalez-Martinez:2007} for He + NH($^3\Sigma$)
collisions with NH in $n=0$ states, shown in Fig.\
\ref{fig-HeNH-a-n0}.

The purpose of the present work is to explore a rather more
complicated case, with significant background inelastic
scattering. Eq.\ \ref{eqaares} still holds, but $a_{\rm res}$ can
be complex and the circle in $S_{00}$ then does not point directly
towards the origin. This behavior is shown in the lower panel of
Fig.\ \ref{fig-s-circles}. The elastic cross section is given by
\begin{equation}
\sigma_{\rm el}(B) = \frac{\pi}{k_0^2} |1-S_{00}(k_0)|^2
, \label{eqsigels}
\end{equation}
so at any value of $B$ it depends on the distance between $S_{00}$
and the point X at $S_{00}=+1$. However, the total inelastic cross
section is given by
\begin{equation}
\sigma_{\rm inel}^{\rm tot}(B) = \frac{\pi}{k_0^2} \left(
1-|S_{00}(k_0)|^2 \right)
, \label{eqsiginels}
\end{equation}
and thus depends on the distance of $S_{00}$ from the unit circle.
If the circle in $S_{00}$ does not point directly towards the
origin, it is clear from Fig.\ \ref{fig-s-circles} that the total
inelastic cross section can show a trough as well as a peak near
resonance. This offers the hope that resonances can be used to
{\em reduce} inelastic rates as well as increase them.

In the general case, the explicit expressions for the real and
imaginary parts of $a(B)$ are \cite{Hutson:res:2007}
\begin{eqnarray}
\alpha(B)=\alpha_{\rm bg} + \frac{\alpha_{\rm res}\left[2(B-B_{\rm
res})/\Gamma_B^{\rm inel}\right] - \beta_{\rm
res}}{\left[2(B-B_{\rm res})/\Gamma_B^{\rm inel}\right]^2+1};\nonumber\\
\beta(B)=\beta_{\rm bg} + \frac{\alpha_{\rm res} + \beta_{\rm
res}\left[2(B-B_{\rm res})/\Gamma_B^{\rm inel}\right]
}{\left[2(B-B_{\rm res})/\Gamma_B^{\rm inel}\right]^2+1},
\label{eqabres}
\end{eqnarray}
where $a(B)=\alpha(B) - {\rm i}\beta(B)$ and similarly for $a_{\rm
res}$ and $a_{\rm bg}$. The peak profiles for the elastic and
total inelastic cross sections are given by combining these with
Eqs.\ \ref{eqsigela} and \ref{eqsiginela}.

\begin{figure}
\includegraphics[width=0.95\linewidth]{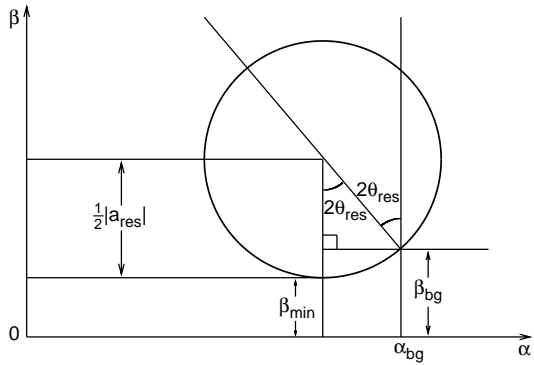}
\caption{The resonant circle in the complex scattering length,
showing the extent to which the imaginary part of $a_{\rm res}$
can reduce inelastic scattering.} \label{fig-a-circle}
\end{figure}

Some useful properties of the scattering length and cross sections
follow from simple geometrical considerations. The complex
scattering length $a(B)$ has a value $a_{\rm bg}$ far from
resonance and describes a circle of radius $|a_{\rm res}|/2$ in
the complex plane as $B$ is tuned across the resonance. If we
write the resonant scattering length as
\begin{equation}
a_{\rm res} =
|a_{\rm res}| \exp\left(2{\rm i}\theta_{\rm res}\right),
\end{equation}
where $\theta_{\rm res} = \phi_0+k_0\alpha_{\rm bg}$, then the
circle is as shown in Fig.\ \ref{fig-a-circle}. The smallest value
achieved by $\beta(B)$ is
\begin{equation}
\beta_{\rm min} = \beta_{\rm bg} - \frac{1}{2}|a_{\rm res}|
(1-\cos 2\theta_{\rm res}),
\end{equation}
which occurs at $B_{\rm min} = B_{\rm res} + x_{\rm
min}\Gamma_B^{\rm inel}/2$ with
\begin{equation}
x_{\rm min}=-(\alpha_{\rm res}/\beta_{\rm res}) -
\left[(\alpha_{\rm res}/\beta_{\rm res})^2+1\right]^{1/2}.
\end{equation}
This defines the smallest value of the total inelastic cross
section through Eq.\ \ref{eqsigela}. Unitarity requires that
$|S_{00}|\le 1$ and $\beta(B)\ge0$, so the limits on the possible
values of $\theta_{\rm res}$ are
\begin{equation}
\cos 2\theta_{\rm res}  \ge \cos 2\theta_{\rm res}^{\rm max} =
1-\frac{2\beta_{\rm bg}}{|a_{\rm res}|}.
\end{equation}
An obvious special case of this is that, if $a_{\rm bg}$ is real,
$\theta_{\rm res}=0$ so that $a_{\rm res}$ is also real.

\begin{figure*}
\includegraphics[width=0.95\linewidth]{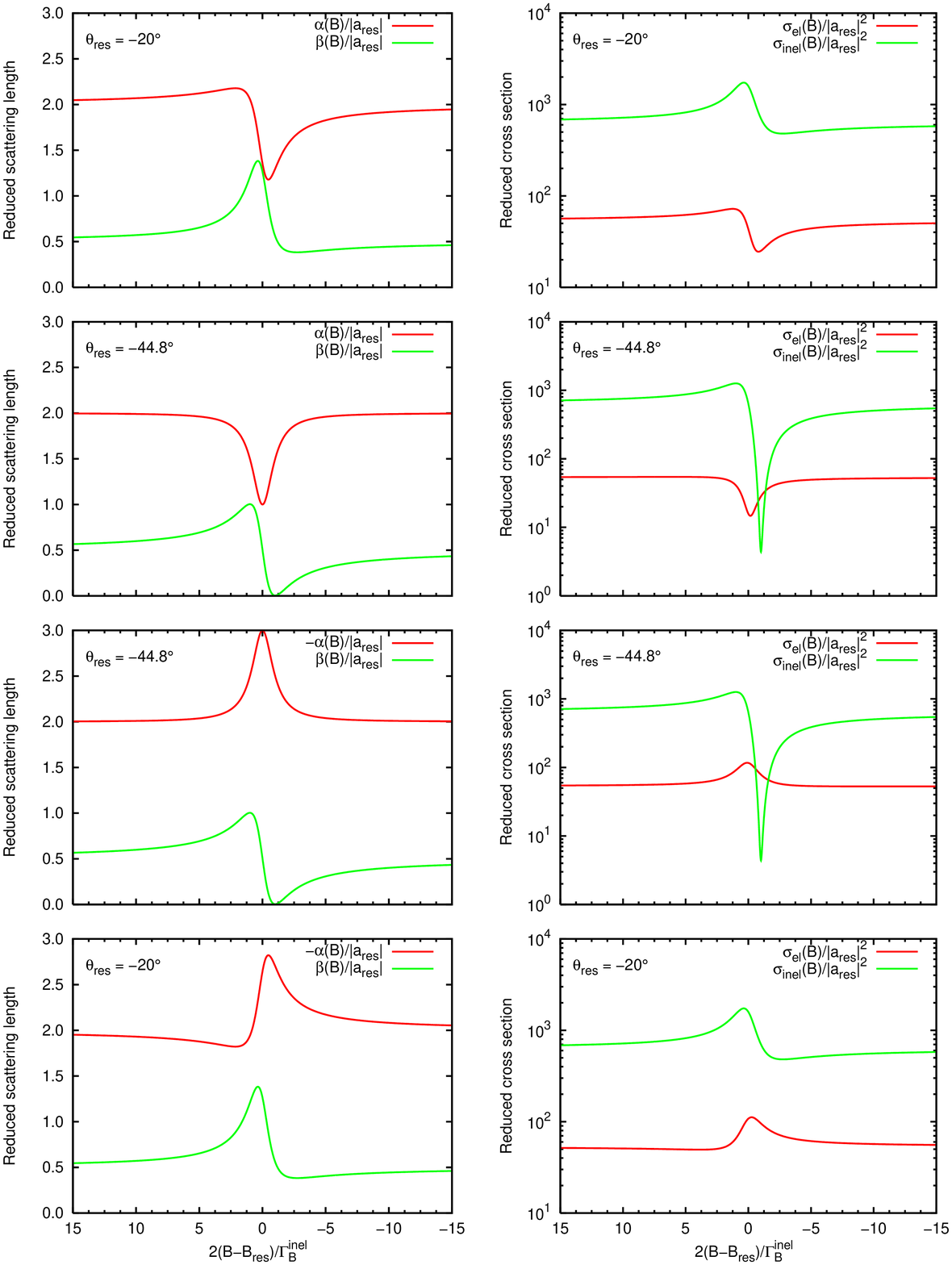}
\caption{(Color online). Some examples of the possible behavior
for a case with moderately strong background inelasticity,
$\beta_{\rm bg}=|a_{\rm res}|/2$ and $|\alpha_{\rm bg}|=2|a_{\rm
res}|$. The left-hand side shows the real and imaginary parts of
the scattering length and the right-hand side shows the elastic
and total inelastic cross sections (with inelastic cross sections
calculated for a wave vector $k_0=10^{-2}|a_{\rm res}|^{-1}$). The
4 panels show (from top to bottom): (i) positive $\alpha_{\rm
bg}$, $\theta_{\rm res}=-20^\circ$; (ii) positive $\alpha_{\rm
bg}$, $\theta_{\rm res}=44^\circ$; (iii) negative $\alpha_{\rm
bg}$, $\theta_{\rm res}=-44^\circ$; (iv) negative $\alpha_{\rm
bg}$, $\theta_{\rm res}=-20^\circ$.} \label{fig-range}
\end{figure*}

Some examples of the possible behavior are illustrated in Fig.\
\ref{fig-range} for a case with moderately strong background
inelasticity, $\beta_{\rm bg}=|a_{\rm res}|/2$ and $|\alpha_{\rm
bg}|=2|a_{\rm res}|$. For these parameters, $\cos\theta_{\rm
res}^{\rm max}=0$. When $\theta_{\rm res}$ is close to its maximum
value, $\beta(B)$ dips close to zero and the total inelastic cross
section shows a trough that can reduce inelastic collision rates
by more than a factor of 10. The elastic cross section also
oscillates, but if $|a_{\rm res}|\ll |\alpha_{\rm bg}|$ the
oscillation is relatively weak and if $\alpha_{\rm bg}$ and
$\beta_{\rm res}$ have opposite signs it is peak-like rather than
trough-like.

\section{Resonances in He + NH ($n=1$)}

We have carried out bound-state and scattering calculations on He
+ NH ($n=1$) in a magnetic field using methods almost identical to
those used previously for $n=0$ \cite{Gonzalez-Martinez:2007}. The
bound-state Schr\"odinger equation was solved using the BOUND
package \cite{Hutson:bound:1993}, as modified to handle magnetic
fields \cite{Gonzalez-Martinez:2007}. In the presence of a
magnetic field, the total angular momentum is no longer a good
quantum number. The calculations are therefore carried out in a
completely decoupled basis set, $|nm_n\rangle |sm_s\rangle
|Lm_L\rangle$, where $s=1$ is the electron spin of NH and $L$ is
the end-over-end rotational angular momentum of He about NH. All
the $m$ quantum numbers represent space-fixed projections on the
axis defined by the magnetic field. The only good quantum numbers
are the parity $(-1)^{n+L+1}$ and the total projection quantum
number $M_{\rm tot} = m_n+m_s+m_L$.

BOUND propagates a set of coupled differential equations outwards
from a point $R_{\rm min}$, deep inside the inner classically
forbidden region, and inwards from a boundary point $R_{\rm max}$
at long range. The two solutions meet at a matching point $R_{\rm
mid}$, and bound-state eigenvalues are found by locating values of
the energy for which the inward and outward solutions can be
matched. The procedure used by BOUND is to seek energies at which
one of the eigenvalues of the log-derivative matching matrix is
zero \cite{Hutson:cpc:1994}.

For true bound states, $R_{\rm max}$ can be placed in the outer
classically forbidden region. However, in the present work we are
dealing with states of He-NH that lie close to the $n=1$ threshold
and are thus more than 30 cm$^{-1}$ above the $n=0$ thresholds.
Since there are open channels, these are actually quasibound
states and can predissociate to form He + NH ($n=0$).
Nevertheless, they can still be located by artificially applying a
{\it bound state} boundary condition at $R_{\rm max}$, and this is
how BOUND is used in the present work.

\begin{figure}
\includegraphics[width=0.98\linewidth]{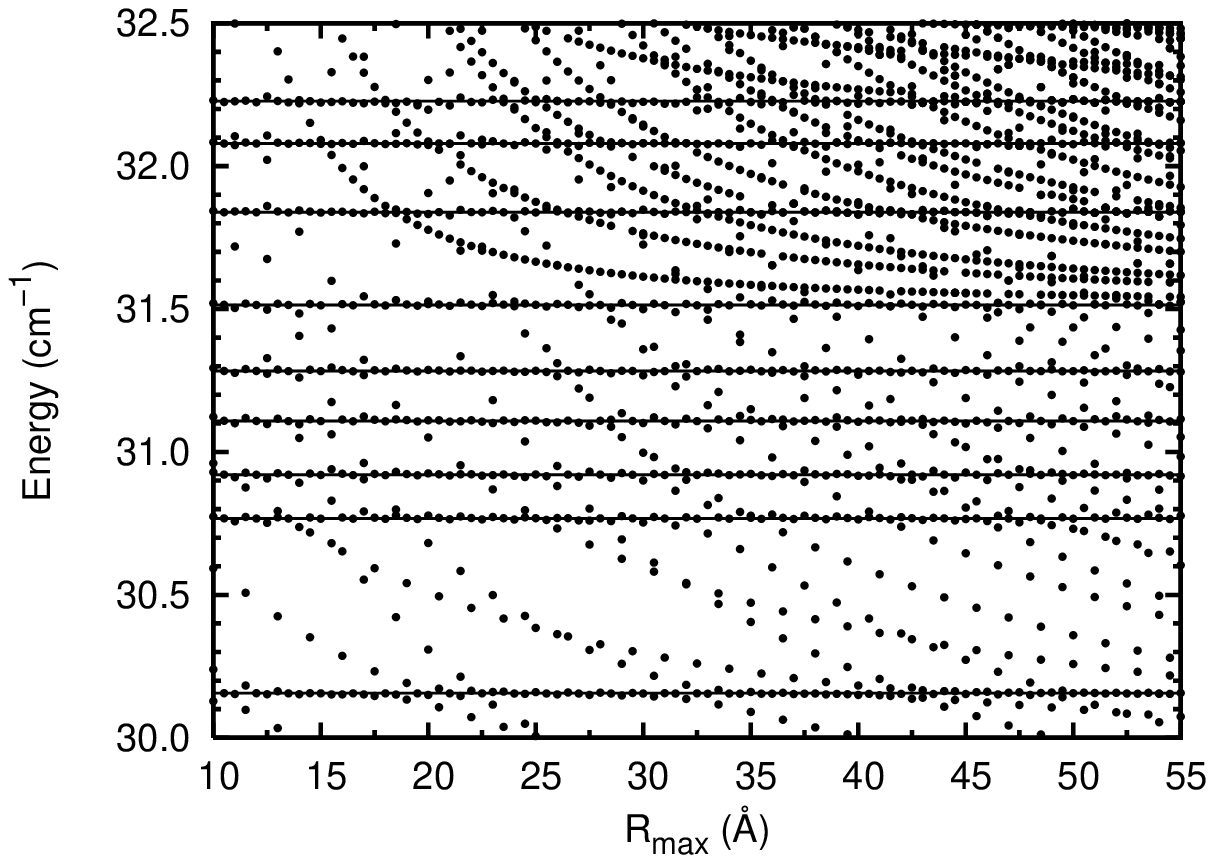}
\includegraphics[width=0.98\linewidth]{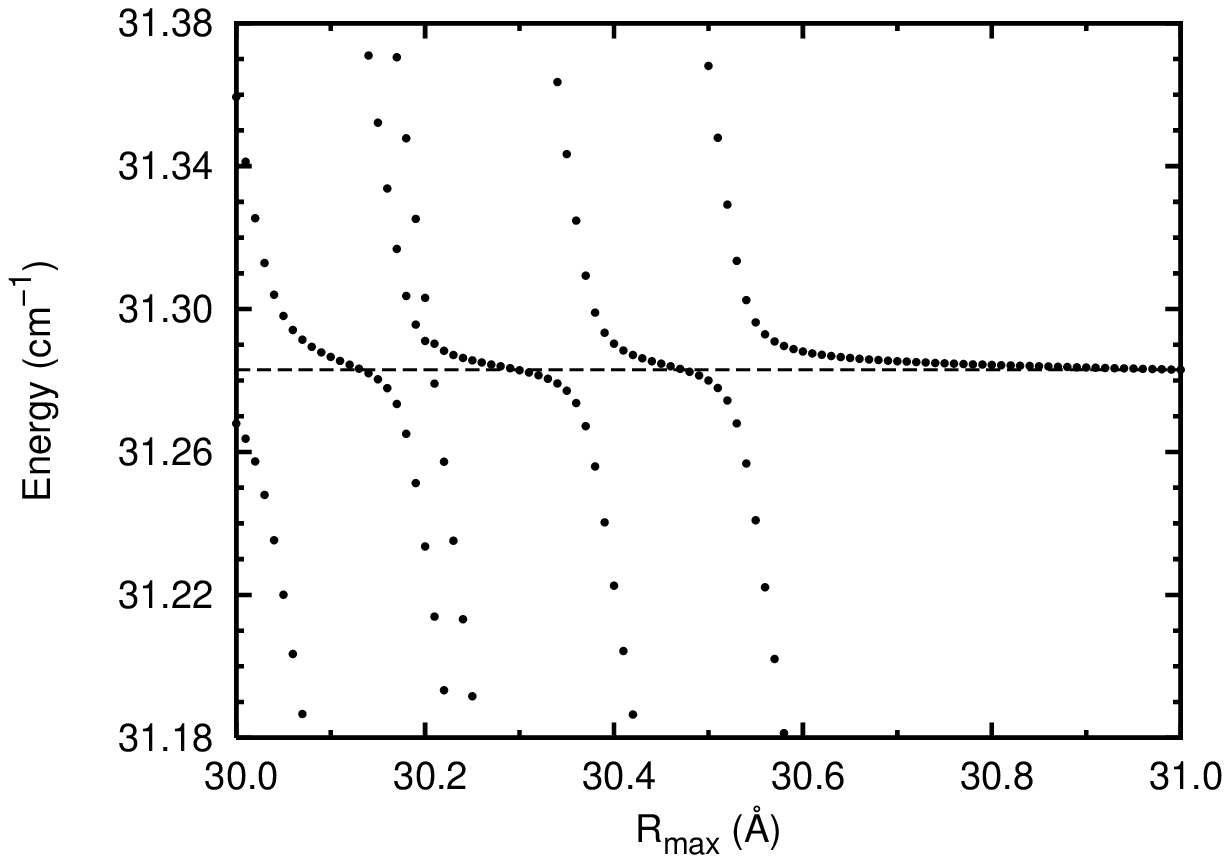}
\caption{The pattern of levels from bound-state calculations on
He-NH near the $n=1$ threshold as a function of $R_{\rm max}$. The
calculations are for $B=4000$ G, even parity, $M_{\rm tot}=0$.
Note the artificial bound states with energies that decrease with
$R_{\rm max}$, crossing and interfering with the physical states
at constant energy. The lower panel shows a more detailed scan
across a small region of $R_{\rm max}$ showing avoided crossings
between real and artificial states.} \label{fig-alllevels}
\end{figure}

Applying a bound-state boundary condition has the side-effect of
box-quantizing the continuum states above both the $n=0$ and $n=1$
thresholds. The resulting {\em artificial} bound states are easily
identified because their energies depend on $R_{\rm max}$.
Difficulties arise only if an artificial bound state lies
accidentally close to the level of interest, in which case the two
states can perturb one another. Fig.\ \ref{fig-alllevels} shows an
example of artificial levels crossing the real levels as a function
of $R_{\rm max}$. For the case of He-NH it was usually possible to
estimate the positions of the physical bound levels to within 0.01
cm$^{-1}$. The perturbations are a measure of the genuine couplings
to the continuum and are comparable to the width of the quasibound
state, so this accuracy is sufficient for use in locating resonance
positions.

\begin{figure}
\includegraphics[width=0.98\linewidth]{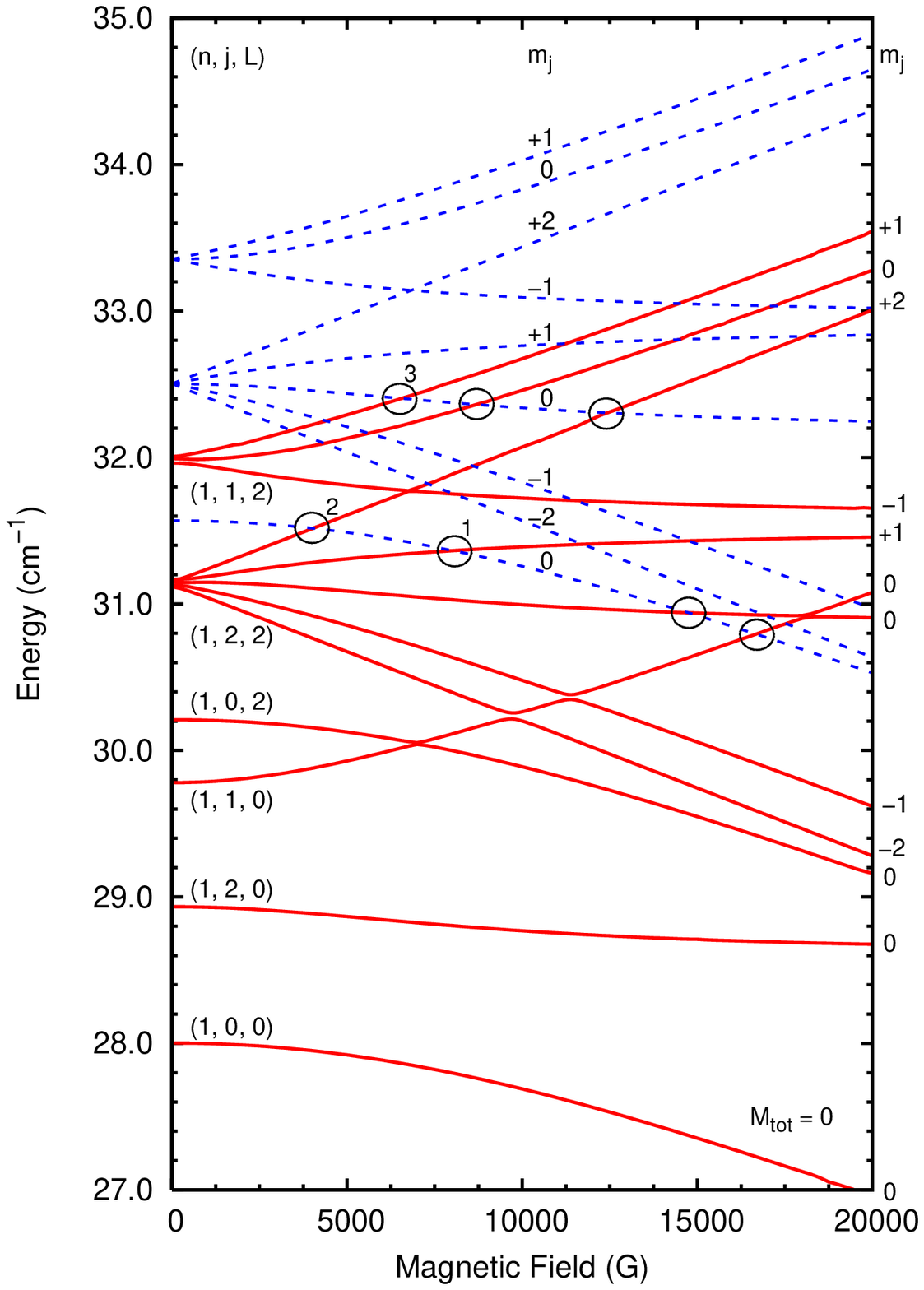}
\caption{(Color online). The pattern of levels from bound-state
calculations on He-NH near the $n=1$ threshold, with artificial
levels removed, as a function of magnetic field $B$. The
calculations are for even parity, $M_{\rm tot}=0$. The circles
show crossings between bound states and thresholds that produce
zero-energy Feshbach resonances.} \label{fig-levs-B}
\end{figure}

Fig.\ \ref{fig-levs-B} shows the quasibound states of He-NH near
the $n=1$ threshold with artificial levels removed. Crossings
between quasibound states and thresholds will produce zero-energy
Feshbach resonances in s-wave scattering if an $L=0$ scattering
channel is permitted by the constraints on parity and $M_{\rm
tot}$. For $M_{\rm tot}=0$ this occurs only for thresholds
corresponding to $m_j=0$, as shown by the circles in Fig.\
\ref{fig-levs-B}.

\begin{figure}
\includegraphics[width=0.98\linewidth]{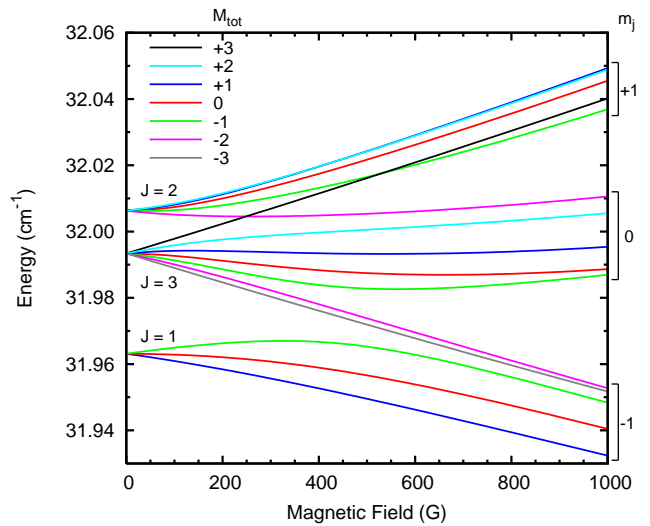}
\caption{(Color online). The pattern of He-NH levels arising from
$(n,j,L)=(1,1,2)$ as a function of magnetic field $B$. The
calculations are for even parity and all allowed values of $M_{\rm
tot}$.} \label{fig-levs-j1}
\end{figure}

Each NH monomer level with $n=1$ is split into 3 components with
$j=0$, 1 and 2 by coupling to the spin $s=1$. The He-NH levels
closest to the $n=1$ thresholds have predominantly $L=2$
character. At zero field the total angular momentum $J$ is a good
quantum number, so for He-NH each $(n,j,L)$ level splits into
$\min(2j+1,2L+1)$ components with different values of $J$. These
splittings are barely visible in Fig.\ \ref{fig-levs-B}, so Fig.\
\ref{fig-levs-j1} shows an expanded view of the levels
corresponding to $(n,j,L)=(1,1,2)$ for all allowed values of
$M_{\rm tot}$. It may be seen that the zero-field levels with
$J=1$, 2 and 3 are split by about 0.04 cm$^{-1}$. When a magnetic
field is applied, each level splits into $2J+1$ components with
different values of $M_{\rm tot}$. The $J$ quantum number remains
a useful label for magnetic fields up to about 200 G, but above
that the levels of different $J$ are strongly mixed. By about 600
G the levels have separated into 3 groups that may be labelled
with an approximate quantum number $m_j$ that takes values $+1$, 0
and $-1$. The levels corresponding to $(n,j,L)$ show similar but
more complex behavior.

Once the crossing points have been located in Fig.\
\ref{fig-levs-B}, the next stage is to carry out scattering
calculations, holding the kinetic energy fixed at a small value
($10^{-6}$ K in the present work) while sweeping the magnetic
field across the resonance. This was done using the MOLSCAT
package \cite{molscat:v14}, as modified to handle magnetic fields
\cite{Gonzalez-Martinez:2007}. MOLSCAT solves the Schr\"odinger
equation by propagating a set of coupled differential equations
outwards from $R_{\rm min}$ to a matching point $R_{\rm max}$ at
long range using basis sets and algorithms very similar to BOUND.
The major difference is that it carries out only outwards
propagation and matches to scattering boundary conditions at
$R_{\rm max}$, so that there is no artificial quantization of the
continuum.

MOLSCAT produces the S matrix and eigenphase sum $\Sigma$ for each
magnetic field $B$ and evaluates the corresponding scattering
length from Eq.\ \ref{eqacomp}. The next step is to fit
$\Sigma(B)$ to Eq.\ \ref{eqbwb} to obtain the resonance position
$B_{\rm res}$ and width $\Gamma_B$. This is done with the RESFIT
package \cite{Hutson:resfit:2007}, which includes a quadratic
polynomial in $B$ for the background term $\Sigma_{\rm bg}$.
RESFIT then proceeds to fit the individual diagonal S-matrix
elements to Eq.\ \ref{eqsii}, holding $B_{\rm res}$ and $\Gamma_B$
fixed at the values obtained from the eigenphase sum. It
represents the magnitude and phase of each $S_{ii}$ with a
quadratic polynomial and produces a complex number $g_{Bi}^2$ that
describes the resonant circle in each S-matrix element. This fit
provides all the parameters required to define the real and
imaginary parts of $a_{\rm bg}$ and $a_{\rm res}$ needed for Eqs.\
\ref{eqaares} and \ref{eqabres}. In practice we use a constant
background term (obtained by evaluating the background polynomials
at $B=B_{\rm res}$) in plotting the results of the formulae below.

\begin{table*}[tb]
\begin{tabular}{crcrcccccc} \cline{1-10} Resonance \quad & $M_{\rm
tot}$ & \quad $B_{\rm res}$ (G) & \quad $\Gamma_B^{\rm inel}$ (G)
& $\Gamma_{B0}$ ($10^{-2}$ G) & $\alpha_{\rm bg}$ (\AA) &
$\beta_{\rm bg}$ ($10^{-3}$ \AA) & $\alpha_{\rm res}$ (\AA)&
$\beta_{\rm res}$
($10^{-2}$ \AA) & $\theta_{\rm res}$  \\

\cline{1-10}

1 & 0 & 8154.71 & $-56.19$ & $-3.938$ & 3.2016 & 4.21 & 2.1735 &
9.08
& $-0.0278$\\

2 & 0 & 4078.47 & $-52.21$ & $-3.208$ & 3.2355 & 5.72 & 1.9057 &
8.57
& $-0.0225$\\

3 & 0 & 6644.15 & $-56.95$ & $-2.224$ & 3.1929 & 4.72 & 1.2119 &
4.76
& $-0.0196$\\

4 & $-2$ & 4250.00 & $-33.64$ & $-2.804$ & 3.1931 & 4.50 & 2.5856
&10.04
& $-0.0194$\\

\cline{1-10}
\end{tabular}
\caption{Parameters of magnetically tuned Feshbach resonances in
He + NH ($n=1$) collisions at $E_{\rm kin}=10^{-6}$ K,
corresponding to $k_0=3.2189 \times 10^{-4}$ \AA$^{-1}$.
Resonances 1 to 3 correspond to the correspondingly numbered
circles in Fig.\ \ref{fig-levs-B}, while resonance 4 is for a
different value of $M_{\rm tot}$.} \label{tabres}
\end{table*}

The results of fitting parameters to several resonances are shown
in Table \ref{tabres}. The first point to notice is that the
resonances are all very wide, with $|\Gamma_B|>30$ G. This
contrasts with the $n=0$ resonances previously characterized
\cite{Gonzalez-Martinez:2007}, which had $|\Gamma_B|<10^{-2}$ G.
The difference arises because the $n=1$ closed channels involved
here are directly coupled to $n=0$ channels by the (weak)
potential anisotropy, whereas the $n=0$ closed channels involved
in our previous work were only indirectly coupled to open channels
by a second-order mechanism involving both the potential
anisotropy and spin-spin coupling. The second point of interest is
that $a_{\rm res}$ has a significant imaginary part in each case,
with $\beta_{\rm res}/\alpha_{\rm res}$ considerably greater than
$\beta_{\rm bg}/\alpha_{\rm bg}$. Because of this, there is
significant asymmetry in the calculated resonant line shapes for
$\alpha(B)$ and $\beta(B)$.

\begin{figure*}
\begin{center}
\includegraphics[width=0.95\linewidth]{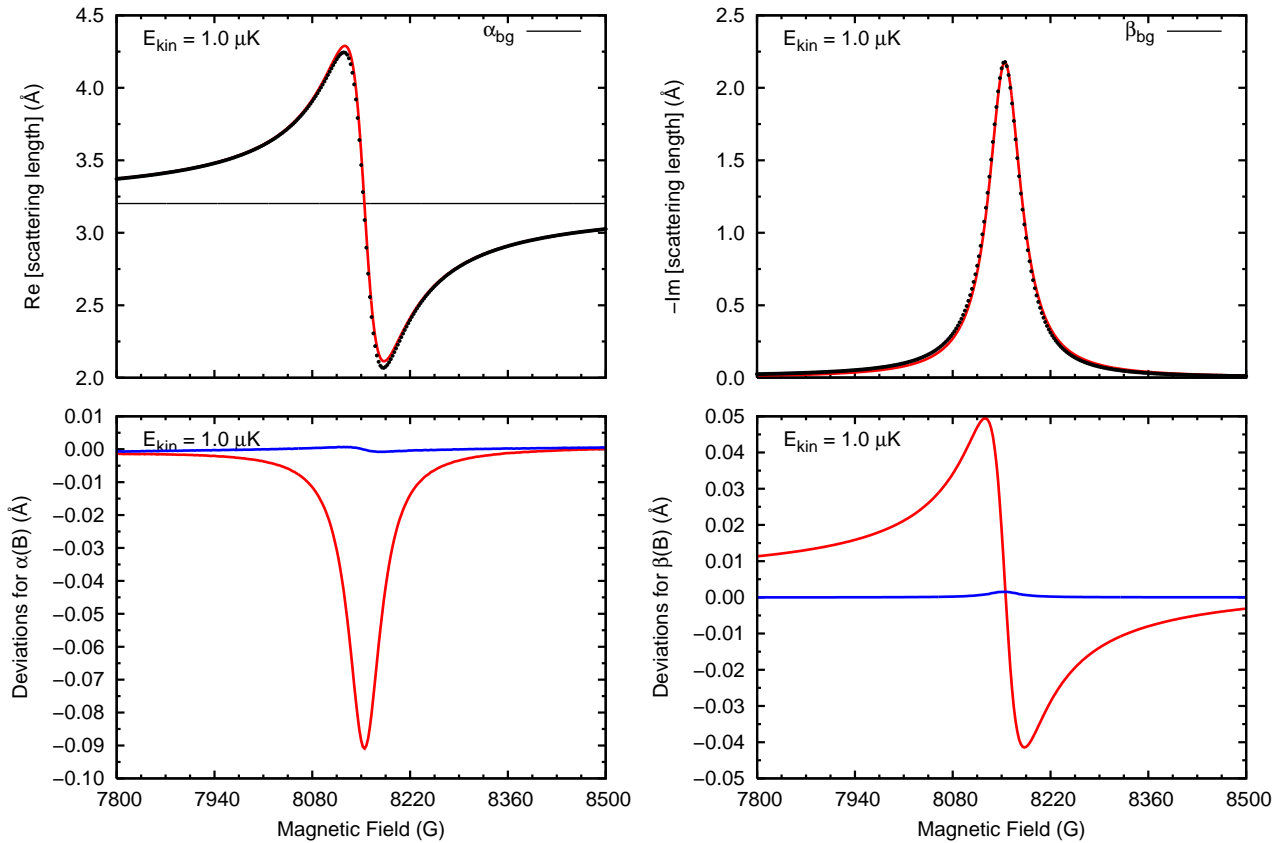}
\caption{(Color online). Upper panels: the real (left) and
imaginary (right) parts of the scattering length for resonance 1
as a function of magnetic field. The black dots show numerical
results from MOLSCAT and the red line shows the result of using
Eq.\ (\ref{eqaares}) with a real value of $a_{\rm res}$. Lower
panel: deviations between the numerical results and the formula
using real (red, Eq.\ \ref{eqaares}) and complex (blue, Eq.\
\ref{eqabres}) values of $a_{\rm res}$.} \label{fig-res1}
\end{center}
\end{figure*}

The real and imaginary scattering lengths for the resonance
labelled 1 in Fig.\ \ref{fig-levs-B} are shown in the upper panels
of Fig.\ \ref{fig-res1}. In each case the black points shows the
numerical results from MOLSCAT and the red line shows the result
of Eq.\ \ref{eqaares} with the assumption that $a_{\rm res}$ is
real (i.e., $a_{\rm res}$ replaced by $|a_{\rm res}|$). It may be
seen that there are significant discrepancies, shown by the red
lines in the two lower panels. By contrast, if $a_{\rm res}$ is
allowed to be complex (Eq.\ \ref{eqabres}), we obtain almost
perfect fits to the numerical results except for the neglect of
the field-dependence of the background scattering length $a_{\rm
bg}$. The result is too close to the points to show usefully in
the upper panels of Fig.\ \ref{fig-res1}, but the difference
between the MOLSCAT results and those given by Eq.\ \ref{eqabres}
are shown as the blue lines in Figs.\ \ref{fig-res1}.

These results verify that Eq.\ \ref{eqabres} gives a correct
account of the the behavior of the scattering length in the
presence of significant (but still small) background inelastic
scattering. The complex nature of $a_{\rm res}$ manifests itself
in a slight tilting of the circles in the S-matrix elements, as
shown in Fig.\ \ref{fig-s-circles}. The tilt is too small to be
shown graphically, but may be seen in the numerical values in
Table \ref{tabres}.

The asymmetries seen in Fig.\ \ref{fig-res1} are much smaller than
the possible asymmetries shown in Fig.\ \ref{fig-range}, but they
nevertheless serve to illustrate the principle that resonant
signatures can be asymmetric and that inelastic cross sections can
show troughs as well as peaks near resonance. In future work we
will investigate systems with stronger background inelasticity in
which more dramatic asymmetries can be expected.

\subsection{Computational details}
The bound and scattering calculations in the present work used a
basis set with $n_{\rm max}=5$ and $L_{\rm max}=5$. The coupled
equations were solved using Johnson's log-derivative algorithm
\cite{Johnson:1973} with $R_{\rm min}=1.7$ \AA\ and a step size of
0.025 \AA. The bound-state calculations used $R_{\rm max}=12$ \AA\
except where stated otherwise, and the scattering calculations
used $R_{\rm max}=100$ \AA.

\section{Conclusions}

We have investigated the behavior of low-energy Feshbach
resonances for a case where there is significant background
inelastic scattering far from resonance. We have located
low-energy Feshbach resonances in s-wave scattering of NH
($^3\Sigma^-$, $n=1$) with He. The resonances are much wider
($\Gamma_B>30$ G) than those found previously for scattering of NH
($n=0$) ($\Gamma_B<10^{-2}$ G) \cite{Gonzalez-Martinez:2007}. We
have investigated how scattering lengths and cross sections vary
as a resonance is tuned across threshold using an external
magnetic field.

The strength of a resonance can be characterized by a {\em
resonant scattering length} $a_{\rm res}$. If only elastic
scattering is possible, $a_{\rm res}$ is infinite and the
scattering length passes through a pole as a bound state crosses
threshold. We have shown previously \cite{Gonzalez-Martinez:2007}
that this behavior is modified in the presence of inelastic
collisions and that in some cases the scattering length shows a
small oscillation rather than a pole.

The key result of the present paper is the demonstration that
$a_{\rm res}$ can be complex rather than real and that this allows
both the real and imaginary parts of the scattering length (and
thus elastic and inelastic cross sections) to show both peaks and
troughs near resonance. We have shown that the real and imaginary
parts of the scattering length follow analytical formulas given
previously \cite{Hutson:res:2007}.

The present work offers the hope that tuning close to a Feshbach
resonance can be used to reduce inelastic collision rates and
thereby allow evaporative or sympathetic cooling in cases where
collisional trap losses would otherwise prevent it. In future work
we will investigate whether additional influences such as electric
fields can be used to {\it control} the maxima and minima and
allow detailed control of collision rates.

\acknowledgments

The authors are grateful to the EPSRC for funding under grant
EP/E039200/1 and to the Royal Society for an International Joint
Project grant which made this collaboration possible.

\bibliography{../../all}

\begin{thebibliography}{44}
\expandafter\ifx\csname natexlab\endcsname\relax\def\natexlab#1{#1}\fi
\expandafter\ifx\csname bibnamefont\endcsname\relax
  \def\bibnamefont#1{#1}\fi
\expandafter\ifx\csname bibfnamefont\endcsname\relax
  \def\bibfnamefont#1{#1}\fi
\expandafter\ifx\csname citenamefont\endcsname\relax
  \def\citenamefont#1{#1}\fi
\expandafter\ifx\csname url\endcsname\relax
  \def\url#1{\texttt{#1}}\fi
\expandafter\ifx\csname urlprefix\endcsname\relax\def\urlprefix{URL }\fi
\providecommand{\bibinfo}[2]{#2}
\providecommand{\eprint}[2][]{\url{#2}}

\bibitem[{\citenamefont{Hutson and Sold\'{a}n}(2006)}]{Hutson:IRPC:2006}
\bibinfo{author}{\bibfnamefont{J.~M.} \bibnamefont{Hutson}} \bibnamefont{and}
  \bibinfo{author}{\bibfnamefont{P.}~\bibnamefont{Sold\'{a}n}},
  \bibinfo{journal}{Int. Rev. Phys. Chem.} \textbf{\bibinfo{volume}{25}},
  \bibinfo{pages}{497} (\bibinfo{year}{2006}).

\bibitem[{\citenamefont{K\"{o}hler et~al.}(2006)\citenamefont{K\"{o}hler,
  Goral, and Julienne}}]{Koehler:RMP:2006}
\bibinfo{author}{\bibfnamefont{T.}~\bibnamefont{K\"{o}hler}},
  \bibinfo{author}{\bibfnamefont{K.}~\bibnamefont{Goral}}, \bibnamefont{and}
  \bibinfo{author}{\bibfnamefont{P.~S.} \bibnamefont{Julienne}},
  \bibinfo{journal}{Rev. Mod. Phys.} \textbf{\bibinfo{volume}{78}},
  \bibinfo{pages}{1311} (\bibinfo{year}{2006}).

\bibitem[{\citenamefont{Roberts et~al.}(2001)\citenamefont{Roberts, Claussen,
  Cornish, Donley, Cornell, and Wieman}}]{Roberts:2001}
\bibinfo{author}{\bibfnamefont{J.~L.} \bibnamefont{Roberts}},
  \bibinfo{author}{\bibfnamefont{N.~R.} \bibnamefont{Claussen}},
  \bibinfo{author}{\bibfnamefont{S.~L.} \bibnamefont{Cornish}},
  \bibinfo{author}{\bibfnamefont{E.~A.} \bibnamefont{Donley}},
  \bibinfo{author}{\bibfnamefont{E.~A.} \bibnamefont{Cornell}},
  \bibnamefont{and} \bibinfo{author}{\bibfnamefont{C.~E.}
  \bibnamefont{Wieman}}, \bibinfo{journal}{Phys. Rev. Lett.}
  \textbf{\bibinfo{volume}{86}}, \bibinfo{pages}{4211} (\bibinfo{year}{2001}).

\bibitem[{\citenamefont{Donley et~al.}(2002)\citenamefont{Donley, Claussen,
  Thompson, and Wieman}}]{Donley:2002}
\bibinfo{author}{\bibfnamefont{E.~A.} \bibnamefont{Donley}},
  \bibinfo{author}{\bibfnamefont{N.~R.} \bibnamefont{Claussen}},
  \bibinfo{author}{\bibfnamefont{S.~T.} \bibnamefont{Thompson}},
  \bibnamefont{and} \bibinfo{author}{\bibfnamefont{C.~E.}
  \bibnamefont{Wieman}}, \bibinfo{journal}{Nature}
  \textbf{\bibinfo{volume}{417}}, \bibinfo{pages}{529} (\bibinfo{year}{2002}).

\bibitem[{\citenamefont{Herbig et~al.}(2003)\citenamefont{Herbig, Kraemer,
  Mark, Weber, Chin, N\"{a}gerl, and Grimm}}]{Herbig:2003}
\bibinfo{author}{\bibfnamefont{J.}~\bibnamefont{Herbig}},
  \bibinfo{author}{\bibfnamefont{T.}~\bibnamefont{Kraemer}},
  \bibinfo{author}{\bibfnamefont{M.}~\bibnamefont{Mark}},
  \bibinfo{author}{\bibfnamefont{T.}~\bibnamefont{Weber}},
  \bibinfo{author}{\bibfnamefont{C.}~\bibnamefont{Chin}},
  \bibinfo{author}{\bibfnamefont{H.~C.} \bibnamefont{N\"{a}gerl}},
  \bibnamefont{and} \bibinfo{author}{\bibfnamefont{R.}~\bibnamefont{Grimm}},
  \bibinfo{journal}{Science} \textbf{\bibinfo{volume}{301}},
  \bibinfo{pages}{1510} (\bibinfo{year}{2003}).

\bibitem[{\citenamefont{Xu et~al.}(2003)\citenamefont{Xu, Mukaiyama,
  Abo-Shaeer, Chin, Miller, and Ketterle}}]{Xu:2003}
\bibinfo{author}{\bibfnamefont{K.}~\bibnamefont{Xu}},
  \bibinfo{author}{\bibfnamefont{T.}~\bibnamefont{Mukaiyama}},
  \bibinfo{author}{\bibfnamefont{J.~R.} \bibnamefont{Abo-Shaeer}},
  \bibinfo{author}{\bibfnamefont{J.~K.} \bibnamefont{Chin}},
  \bibinfo{author}{\bibfnamefont{D.~E.} \bibnamefont{Miller}},
  \bibnamefont{and} \bibinfo{author}{\bibfnamefont{W.}~\bibnamefont{Ketterle}},
  \bibinfo{journal}{Phys. Rev. Lett.} \textbf{\bibinfo{volume}{91}},
  \bibinfo{pages}{210402} (\bibinfo{year}{2003}).

\bibitem[{\citenamefont{D\"urr et~al.}(2004)\citenamefont{D\"urr, Volz, Marte,
  and Rempe}}]{Durr:mol87Rb:2004}
\bibinfo{author}{\bibfnamefont{S.}~\bibnamefont{D\"urr}},
  \bibinfo{author}{\bibfnamefont{T.}~\bibnamefont{Volz}},
  \bibinfo{author}{\bibfnamefont{A.}~\bibnamefont{Marte}}, \bibnamefont{and}
  \bibinfo{author}{\bibfnamefont{G.}~\bibnamefont{Rempe}},
  \bibinfo{journal}{Phys. Rev. Lett.} \textbf{\bibinfo{volume}{92}},
  \bibinfo{pages}{020406} (\bibinfo{year}{2004}).

\bibitem[{\citenamefont{Regal et~al.}(2003)\citenamefont{Regal, Ticknor, Bohn,
  and Jin}}]{Regal:40K2:2003}
\bibinfo{author}{\bibfnamefont{C.~A.} \bibnamefont{Regal}},
  \bibinfo{author}{\bibfnamefont{C.}~\bibnamefont{Ticknor}},
  \bibinfo{author}{\bibfnamefont{J.~L.} \bibnamefont{Bohn}}, \bibnamefont{and}
  \bibinfo{author}{\bibfnamefont{D.~S.} \bibnamefont{Jin}},
  \bibinfo{journal}{Nature} \textbf{\bibinfo{volume}{424}}, \bibinfo{pages}{47}
  (\bibinfo{year}{2003}).

\bibitem[{\citenamefont{Strecker et~al.}(2003)\citenamefont{Strecker,
  Partridge, and Hulet}}]{Strecker:2003}
\bibinfo{author}{\bibfnamefont{K.~E.} \bibnamefont{Strecker}},
  \bibinfo{author}{\bibfnamefont{G.~B.} \bibnamefont{Partridge}},
  \bibnamefont{and} \bibinfo{author}{\bibfnamefont{R.~G.} \bibnamefont{Hulet}},
  \bibinfo{journal}{Phys. Rev. Lett.} \textbf{\bibinfo{volume}{91}},
  \bibinfo{pages}{080406} (\bibinfo{year}{2003}).

\bibitem[{\citenamefont{Cubizolles et~al.}(2003)\citenamefont{Cubizolles,
  Bourdel, Kokkelmans, Shlyapnikov, and Salomon}}]{Cubizolles:2003}
\bibinfo{author}{\bibfnamefont{J.}~\bibnamefont{Cubizolles}},
  \bibinfo{author}{\bibfnamefont{T.}~\bibnamefont{Bourdel}},
  \bibinfo{author}{\bibfnamefont{S.~J. J. M.~F.} \bibnamefont{Kokkelmans}},
  \bibinfo{author}{\bibfnamefont{G.~V.} \bibnamefont{Shlyapnikov}},
  \bibnamefont{and} \bibinfo{author}{\bibfnamefont{C.}~\bibnamefont{Salomon}},
  \bibinfo{journal}{Phys. Rev. Lett.} \textbf{\bibinfo{volume}{91}},
  \bibinfo{pages}{240401} (\bibinfo{year}{2003}).

\bibitem[{\citenamefont{Jochim et~al.}(2003{\natexlab{a}})\citenamefont{Jochim,
  Bartenstein, Altmeyer, Hendl, Chin, Denschlag, and
  Grimm}}]{Jochim:Li2pure:2003}
\bibinfo{author}{\bibfnamefont{S.}~\bibnamefont{Jochim}},
  \bibinfo{author}{\bibfnamefont{M.}~\bibnamefont{Bartenstein}},
  \bibinfo{author}{\bibfnamefont{A.}~\bibnamefont{Altmeyer}},
  \bibinfo{author}{\bibfnamefont{G.}~\bibnamefont{Hendl}},
  \bibinfo{author}{\bibfnamefont{C.}~\bibnamefont{Chin}},
  \bibinfo{author}{\bibfnamefont{J.~H.} \bibnamefont{Denschlag}},
  \bibnamefont{and} \bibinfo{author}{\bibfnamefont{R.}~\bibnamefont{Grimm}},
  \bibinfo{journal}{Phys. Rev. Lett.} \textbf{\bibinfo{volume}{91}},
  \bibinfo{pages}{240402} (\bibinfo{year}{2003}{\natexlab{a}}).

\bibitem[{\citenamefont{Jochim et~al.}(2003{\natexlab{b}})\citenamefont{Jochim,
  Bartenstein, Altmeyer, Hendl, Riedl, Chin, Denschlag, and
  Grimm}}]{Jochim:Li2BEC:2003}
\bibinfo{author}{\bibfnamefont{S.}~\bibnamefont{Jochim}},
  \bibinfo{author}{\bibfnamefont{M.}~\bibnamefont{Bartenstein}},
  \bibinfo{author}{\bibfnamefont{A.}~\bibnamefont{Altmeyer}},
  \bibinfo{author}{\bibfnamefont{G.}~\bibnamefont{Hendl}},
  \bibinfo{author}{\bibfnamefont{S.}~\bibnamefont{Riedl}},
  \bibinfo{author}{\bibfnamefont{C.}~\bibnamefont{Chin}},
  \bibinfo{author}{\bibfnamefont{J.~H.} \bibnamefont{Denschlag}},
  \bibnamefont{and} \bibinfo{author}{\bibfnamefont{R.}~\bibnamefont{Grimm}},
  \bibinfo{journal}{Science} \textbf{\bibinfo{volume}{302}},
  \bibinfo{pages}{2101} (\bibinfo{year}{2003}{\natexlab{b}}).

\bibitem[{\citenamefont{Zwierlein et~al.}(2003)\citenamefont{Zwierlein, Stan,
  Schunck, Raupach, Gupta, Hadzibabic, and Ketterle}}]{Zwierlein:2003}
\bibinfo{author}{\bibfnamefont{M.~W.} \bibnamefont{Zwierlein}},
  \bibinfo{author}{\bibfnamefont{C.~A.} \bibnamefont{Stan}},
  \bibinfo{author}{\bibfnamefont{C.~H.} \bibnamefont{Schunck}},
  \bibinfo{author}{\bibfnamefont{S.~M.~F.} \bibnamefont{Raupach}},
  \bibinfo{author}{\bibfnamefont{S.}~\bibnamefont{Gupta}},
  \bibinfo{author}{\bibfnamefont{Z.}~\bibnamefont{Hadzibabic}},
  \bibnamefont{and} \bibinfo{author}{\bibfnamefont{W.}~\bibnamefont{Ketterle}},
  \bibinfo{journal}{Phys. Rev. Lett.} \textbf{\bibinfo{volume}{91}},
  \bibinfo{pages}{250401} (\bibinfo{year}{2003}).

\bibitem[{\citenamefont{Greiner et~al.}(2003)\citenamefont{Greiner, Regal, and
  Jin}}]{Greiner:2003}
\bibinfo{author}{\bibfnamefont{M.}~\bibnamefont{Greiner}},
  \bibinfo{author}{\bibfnamefont{C.~A.} \bibnamefont{Regal}}, \bibnamefont{and}
  \bibinfo{author}{\bibfnamefont{D.~S.} \bibnamefont{Jin}},
  \bibinfo{journal}{Nature} \textbf{\bibinfo{volume}{426}},
  \bibinfo{pages}{537} (\bibinfo{year}{2003}).

\bibitem[{\citenamefont{Kraemer et~al.}(2006)\citenamefont{Kraemer, Mark,
  Waldburger, Danzl, Chin, Engeser, Lange, Pilch, Jaakkola, N\"{a}gerl
  et~al.}}]{Kraemer:2006}
\bibinfo{author}{\bibfnamefont{T.}~\bibnamefont{Kraemer}},
  \bibinfo{author}{\bibfnamefont{M.}~\bibnamefont{Mark}},
  \bibinfo{author}{\bibfnamefont{P.}~\bibnamefont{Waldburger}},
  \bibinfo{author}{\bibfnamefont{J.~G.} \bibnamefont{Danzl}},
  \bibinfo{author}{\bibfnamefont{C.}~\bibnamefont{Chin}},
  \bibinfo{author}{\bibfnamefont{B.}~\bibnamefont{Engeser}},
  \bibinfo{author}{\bibfnamefont{A.~D.} \bibnamefont{Lange}},
  \bibinfo{author}{\bibfnamefont{K.}~\bibnamefont{Pilch}},
  \bibinfo{author}{\bibfnamefont{A.}~\bibnamefont{Jaakkola}},
  \bibinfo{author}{\bibfnamefont{H.~C.} \bibnamefont{N\"{a}gerl}},
  \bibnamefont{et~al.}, \bibinfo{journal}{Nature}
  \textbf{\bibinfo{volume}{440}}, \bibinfo{pages}{315} (\bibinfo{year}{2006}).

\bibitem[{\citenamefont{Chin et~al.}(2005)\citenamefont{Chin, Kraemer, Mark,
  Herbig, Waldburger, N\"{a}gerl, and Grimm}}]{Chin:2005}
\bibinfo{author}{\bibfnamefont{C.}~\bibnamefont{Chin}},
  \bibinfo{author}{\bibfnamefont{T.}~\bibnamefont{Kraemer}},
  \bibinfo{author}{\bibfnamefont{M.}~\bibnamefont{Mark}},
  \bibinfo{author}{\bibfnamefont{J.}~\bibnamefont{Herbig}},
  \bibinfo{author}{\bibfnamefont{P.}~\bibnamefont{Waldburger}},
  \bibinfo{author}{\bibfnamefont{H.~C.} \bibnamefont{N\"{a}gerl}},
  \bibnamefont{and} \bibinfo{author}{\bibfnamefont{R.}~\bibnamefont{Grimm}},
  \bibinfo{journal}{Phys. Rev. Lett.} \textbf{\bibinfo{volume}{94}},
  \bibinfo{pages}{123201} (\bibinfo{year}{2005}).

\bibitem[{\citenamefont{Mark et~al.}(2007{\natexlab{a}})\citenamefont{Mark,
  Kraemer, Waldburger, Herbig, Chin, Naegerl, and Grimm}}]{Mark:stuck:2007}
\bibinfo{author}{\bibfnamefont{M.}~\bibnamefont{Mark}},
  \bibinfo{author}{\bibfnamefont{T.}~\bibnamefont{Kraemer}},
  \bibinfo{author}{\bibfnamefont{P.}~\bibnamefont{Waldburger}},
  \bibinfo{author}{\bibfnamefont{J.}~\bibnamefont{Herbig}},
  \bibinfo{author}{\bibfnamefont{C.}~\bibnamefont{Chin}},
  \bibinfo{author}{\bibfnamefont{H.-C.} \bibnamefont{Naegerl}},
  \bibnamefont{and} \bibinfo{author}{\bibfnamefont{R.}~\bibnamefont{Grimm}},
  \bibinfo{journal}{arXiv:cond-mat/0704.0653}
  (\bibinfo{year}{2007}{\natexlab{a}}).

\bibitem[{\citenamefont{Mark et~al.}(2007{\natexlab{b}})\citenamefont{Mark,
  Ferlaino, Knoop, Kraemer, Chin, Naegerl, and Grimm}}]{Mark:spect:2007}
\bibinfo{author}{\bibfnamefont{M.}~\bibnamefont{Mark}},
  \bibinfo{author}{\bibfnamefont{F.}~\bibnamefont{Ferlaino}},
  \bibinfo{author}{\bibfnamefont{S.}~\bibnamefont{Knoop}},
  \bibinfo{author}{\bibfnamefont{T.}~\bibnamefont{Kraemer}},
  \bibinfo{author}{\bibfnamefont{C.}~\bibnamefont{Chin}},
  \bibinfo{author}{\bibfnamefont{H.-C.} \bibnamefont{Naegerl}},
  \bibnamefont{and} \bibinfo{author}{\bibfnamefont{R.}~\bibnamefont{Grimm}},
  \bibinfo{journal}{arXiv:cond-mat/0706.1041}
  (\bibinfo{year}{2007}{\natexlab{b}}).

\bibitem[{\citenamefont{Pethick and Smith}(2002)}]{Pethick:2002}
\bibinfo{author}{\bibfnamefont{C.~J.} \bibnamefont{Pethick}} \bibnamefont{and}
  \bibinfo{author}{\bibfnamefont{H.}~\bibnamefont{Smith}},
  \emph{\bibinfo{title}{Bose-Einstein Condensation in Dilute Gases}}
  (\bibinfo{publisher}{Cambridge University Press}, \bibinfo{year}{2002}).

\bibitem[{\citenamefont{Moerdijk et~al.}(1995)\citenamefont{Moerdijk, Verhaar,
  and Axelsson}}]{Moerdijk:1995}
\bibinfo{author}{\bibfnamefont{A.~J.} \bibnamefont{Moerdijk}},
  \bibinfo{author}{\bibfnamefont{B.~J.} \bibnamefont{Verhaar}},
  \bibnamefont{and} \bibinfo{author}{\bibfnamefont{A.}~\bibnamefont{Axelsson}},
  \bibinfo{journal}{Phys. Rev. A} \textbf{\bibinfo{volume}{51}},
  \bibinfo{pages}{4852} (\bibinfo{year}{1995}).

\bibitem[{\citenamefont{Weinstein et~al.}(1998)\citenamefont{Weinstein,
  deCarvalho, Guillet, Friedrich, and Doyle}}]{Weinstein:CaH:1998}
\bibinfo{author}{\bibfnamefont{J.~D.} \bibnamefont{Weinstein}},
  \bibinfo{author}{\bibfnamefont{R.}~\bibnamefont{deCarvalho}},
  \bibinfo{author}{\bibfnamefont{T.}~\bibnamefont{Guillet}},
  \bibinfo{author}{\bibfnamefont{B.}~\bibnamefont{Friedrich}},
  \bibnamefont{and} \bibinfo{author}{\bibfnamefont{J.~M.} \bibnamefont{Doyle}},
  \bibinfo{journal}{Nature} \textbf{\bibinfo{volume}{395}},
  \bibinfo{pages}{148} (\bibinfo{year}{1998}).

\bibitem[{\citenamefont{Egorov et~al.}(2004)\citenamefont{Egorov, Campbell,
  Friedrich, Maxwell, Tsikata, van Buuren, and Doyle}}]{Egorov:2004}
\bibinfo{author}{\bibfnamefont{D.}~\bibnamefont{Egorov}},
  \bibinfo{author}{\bibfnamefont{W.~C.} \bibnamefont{Campbell}},
  \bibinfo{author}{\bibfnamefont{B.}~\bibnamefont{Friedrich}},
  \bibinfo{author}{\bibfnamefont{S.~E.} \bibnamefont{Maxwell}},
  \bibinfo{author}{\bibfnamefont{E.}~\bibnamefont{Tsikata}},
  \bibinfo{author}{\bibfnamefont{L.~D.} \bibnamefont{van Buuren}},
  \bibnamefont{and} \bibinfo{author}{\bibfnamefont{J.~M.} \bibnamefont{Doyle}},
  \bibinfo{journal}{Eur. Phys. J. D} \textbf{\bibinfo{volume}{31}},
  \bibinfo{pages}{307} (\bibinfo{year}{2004}).

\bibitem[{\citenamefont{Campbell et~al.}(2007)\citenamefont{Campbell, Tsikata,
  van Buuren, Lu, and Doyle}}]{Campbell:2007}
\bibinfo{author}{\bibfnamefont{W.~C.} \bibnamefont{Campbell}},
  \bibinfo{author}{\bibfnamefont{E.}~\bibnamefont{Tsikata}},
  \bibinfo{author}{\bibfnamefont{L.}~\bibnamefont{van Buuren}},
  \bibinfo{author}{\bibfnamefont{H.-I.} \bibnamefont{Lu}}, \bibnamefont{and}
  \bibinfo{author}{\bibfnamefont{J.~M.} \bibnamefont{Doyle}},
  \bibinfo{journal}{arXiv:physics/0702071}  (\bibinfo{year}{2007}).

\bibitem[{\citenamefont{Bethlem and Meijer}(2003)}]{Bethlem:IRPC:2003}
\bibinfo{author}{\bibfnamefont{H.~L.} \bibnamefont{Bethlem}} \bibnamefont{and}
  \bibinfo{author}{\bibfnamefont{G.}~\bibnamefont{Meijer}},
  \bibinfo{journal}{Int. Rev. Phys. Chem.} \textbf{\bibinfo{volume}{22}},
  \bibinfo{pages}{73} (\bibinfo{year}{2003}).

\bibitem[{\citenamefont{Bethlem et~al.}(2006)\citenamefont{Bethlem, Tarbutt,
  K\"{u}pper, Carty, Wohlfart, Hinds, and Meijer}}]{Bethlem:2006}
\bibinfo{author}{\bibfnamefont{H.~L.} \bibnamefont{Bethlem}},
  \bibinfo{author}{\bibfnamefont{M.~R.} \bibnamefont{Tarbutt}},
  \bibinfo{author}{\bibfnamefont{J.}~\bibnamefont{K\"{u}pper}},
  \bibinfo{author}{\bibfnamefont{D.}~\bibnamefont{Carty}},
  \bibinfo{author}{\bibfnamefont{K.}~\bibnamefont{Wohlfart}},
  \bibinfo{author}{\bibfnamefont{E.~A.} \bibnamefont{Hinds}}, \bibnamefont{and}
  \bibinfo{author}{\bibfnamefont{G.}~\bibnamefont{Meijer}},
  \bibinfo{journal}{J. Phys. B -- At. Mol. Opt. Phys.}
  \textbf{\bibinfo{volume}{39}}, \bibinfo{pages}{R263} (\bibinfo{year}{2006}).

\bibitem[{\citenamefont{Domokos and Ritsch}(2002)}]{Domokos:2002}
\bibinfo{author}{\bibfnamefont{P.}~\bibnamefont{Domokos}} \bibnamefont{and}
  \bibinfo{author}{\bibfnamefont{H.}~\bibnamefont{Ritsch}},
  \bibinfo{journal}{Phys. Rev. Lett.} \textbf{\bibinfo{volume}{89}},
  \bibinfo{pages}{253003} (\bibinfo{year}{2002}).

\bibitem[{\citenamefont{Chan et~al.}(2003)\citenamefont{Chan, Black, and
  Vuletic}}]{Chan:2003}
\bibinfo{author}{\bibfnamefont{H.~W.} \bibnamefont{Chan}},
  \bibinfo{author}{\bibfnamefont{A.~T.} \bibnamefont{Black}}, \bibnamefont{and}
  \bibinfo{author}{\bibfnamefont{V.}~\bibnamefont{Vuletic}},
  \bibinfo{journal}{Phys. Rev. Lett.} \textbf{\bibinfo{volume}{90}},
  \bibinfo{pages}{063003} (\bibinfo{year}{2003}).

\bibitem[{\citenamefont{Morigi et~al.}(2007)\citenamefont{Morigi, Pinkse,
  Kowalewski, and de~Vivie-Riedle}}]{Morigi:2007}
\bibinfo{author}{\bibfnamefont{G.}~\bibnamefont{Morigi}},
  \bibinfo{author}{\bibfnamefont{P.~W.~H.} \bibnamefont{Pinkse}},
  \bibinfo{author}{\bibfnamefont{M.}~\bibnamefont{Kowalewski}},
  \bibnamefont{and}
  \bibinfo{author}{\bibfnamefont{R.}~\bibnamefont{de~Vivie-Riedle}},
  \bibinfo{journal}{arXiv:quant-ph/0703157}  (\bibinfo{year}{2007}).

\bibitem[{\citenamefont{Gonz\'{a}lez-Mart\'{\i}nez and
  Hutson}(2007)}]{Gonzalez-Martinez:2007}
\bibinfo{author}{\bibfnamefont{M.~L.} \bibnamefont{Gonz\'{a}lez-Mart\'{\i}nez}}
  \bibnamefont{and} \bibinfo{author}{\bibfnamefont{J.~M.}
  \bibnamefont{Hutson}}, \bibinfo{journal}{Phys. Rev. A}
  \textbf{\bibinfo{volume}{75}}, \bibinfo{pages}{022702}
  (\bibinfo{year}{2007}).

\bibitem[{\citenamefont{Hutson}(1993)}]{Hutson:bound:1993}
\bibinfo{author}{\bibfnamefont{J.~M.} \bibnamefont{Hutson}},
  \emph{\bibinfo{title}{Bound computer program, version 5}},
  \bibinfo{howpublished}{distributed by Collaborative Computational Project
  No.\ 6 of the UK Engineering and Physical Sciences Research Council}
  (\bibinfo{year}{1993}).

\bibitem[{\citenamefont{Hutson and Green}(1994)}]{molscat:v14}
\bibinfo{author}{\bibfnamefont{J.~M.} \bibnamefont{Hutson}} \bibnamefont{and}
  \bibinfo{author}{\bibfnamefont{S.}~\bibnamefont{Green}},
  \emph{\bibinfo{title}{Molscat computer program, version 14}},
  \bibinfo{howpublished}{distributed by Collaborative Computational Project
  No.\ 6 of the UK Engineering and Physical Sciences Research Council}
  (\bibinfo{year}{1994}).

\bibitem[{\citenamefont{Hutson}(2007{\natexlab{a}})}]{Hutson:res:2007}
\bibinfo{author}{\bibfnamefont{J.~M.} \bibnamefont{Hutson}},
  \bibinfo{journal}{New J. Phys.} \textbf{\bibinfo{volume}{9}},
  \bibinfo{pages}{152} (\bibinfo{year}{2007}{\natexlab{a}}),
  \bibinfo{note}{note that there is a typographical error in Eq.\ (22) of this
  paper: the last term on the right-hand side should read $-\beta_{\rm res}$
  instead of $+\beta_{\rm res}$.}

\bibitem[{\citenamefont{Mott and Massey}(1965)}]{Mott:p380:1965}
\bibinfo{author}{\bibfnamefont{N.~F.} \bibnamefont{Mott}} \bibnamefont{and}
  \bibinfo{author}{\bibfnamefont{H.~S.~W.} \bibnamefont{Massey}},
  \emph{\bibinfo{title}{The Theory of Atomic Collisions}}
  (\bibinfo{publisher}{Clarendon Press, Oxford}, \bibinfo{year}{1965}),
  \bibinfo{edition}{3rd} ed.

\bibitem[{\citenamefont{Bohn and Julienne}(1997)}]{Bohn:1997}
\bibinfo{author}{\bibfnamefont{J.~L.} \bibnamefont{Bohn}} \bibnamefont{and}
  \bibinfo{author}{\bibfnamefont{P.~S.} \bibnamefont{Julienne}},
  \bibinfo{journal}{Phys. Rev. A} \textbf{\bibinfo{volume}{56}},
  \bibinfo{pages}{1486} (\bibinfo{year}{1997}).

\bibitem[{\citenamefont{Balakrishnan et~al.}(1997)\citenamefont{Balakrishnan,
  Kharchenko, Forrey, and Dalgarno}}]{Balakrishnan:scat-len:1997}
\bibinfo{author}{\bibfnamefont{N.}~\bibnamefont{Balakrishnan}},
  \bibinfo{author}{\bibfnamefont{V.}~\bibnamefont{Kharchenko}},
  \bibinfo{author}{\bibfnamefont{R.~C.} \bibnamefont{Forrey}},
  \bibnamefont{and} \bibinfo{author}{\bibfnamefont{A.}~\bibnamefont{Dalgarno}},
  \bibinfo{journal}{Chem. Phys. Lett.} \textbf{\bibinfo{volume}{280}},
  \bibinfo{pages}{5} (\bibinfo{year}{1997}).

\bibitem[{\citenamefont{Cvita\v{s} et~al.}(2007)\citenamefont{Cvita\v{s},
  Sold\'{a}n, Hutson, Honvault, and Launay}}]{Cvitas:li3:2007}
\bibinfo{author}{\bibfnamefont{M.~T.} \bibnamefont{Cvita\v{s}}},
  \bibinfo{author}{\bibfnamefont{P.}~\bibnamefont{Sold\'{a}n}},
  \bibinfo{author}{\bibfnamefont{J.~M.} \bibnamefont{Hutson}},
  \bibinfo{author}{\bibfnamefont{P.}~\bibnamefont{Honvault}}, \bibnamefont{and}
  \bibinfo{author}{\bibfnamefont{J.~M.} \bibnamefont{Launay}},
  \bibinfo{journal}{arXiv:physics/0703136}  (\bibinfo{year}{2007}).

\bibitem[{\citenamefont{Hazi}(1979)}]{Hazi:1979}
\bibinfo{author}{\bibfnamefont{A.~U.} \bibnamefont{Hazi}},
  \bibinfo{journal}{Phys. Rev. A} \textbf{\bibinfo{volume}{19}},
  \bibinfo{pages}{920} (\bibinfo{year}{1979}).

\bibitem[{\citenamefont{Ashton et~al.}(1983)\citenamefont{Ashton, Child, and
  Hutson}}]{Ashton:1983}
\bibinfo{author}{\bibfnamefont{C.~J.} \bibnamefont{Ashton}},
  \bibinfo{author}{\bibfnamefont{M.~S.} \bibnamefont{Child}}, \bibnamefont{and}
  \bibinfo{author}{\bibfnamefont{J.~M.} \bibnamefont{Hutson}},
  \bibinfo{journal}{J. Chem. Phys.} \textbf{\bibinfo{volume}{78}},
  \bibinfo{pages}{4025} (\bibinfo{year}{1983}).

\bibitem[{\citenamefont{Brenig and Haag}(1959)}]{Brenig:1959}
\bibinfo{author}{\bibfnamefont{W.}~\bibnamefont{Brenig}} \bibnamefont{and}
  \bibinfo{author}{\bibfnamefont{R.}~\bibnamefont{Haag}},
  \bibinfo{journal}{Fortschr. Phys.} \textbf{\bibinfo{volume}{7}},
  \bibinfo{pages}{183} (\bibinfo{year}{1959}).

\bibitem[{\citenamefont{Taylor}(1972)}]{Taylor:p411:1972}
\bibinfo{author}{\bibfnamefont{J.~R.} \bibnamefont{Taylor}},
  \emph{\bibinfo{title}{Scattering Theory: The Quantum Theory of
  Nonrelativistic Collisions}} (\bibinfo{publisher}{Wiley, New York},
  \bibinfo{year}{1972}).

\bibitem[{\citenamefont{Gazdy and Bowman}(1987)}]{Gazdy:1987}
\bibinfo{author}{\bibfnamefont{B.}~\bibnamefont{Gazdy}} \bibnamefont{and}
  \bibinfo{author}{\bibfnamefont{J.~M.} \bibnamefont{Bowman}},
  \bibinfo{journal}{Phys. Rev. Lett.} \textbf{\bibinfo{volume}{59}},
  \bibinfo{pages}{3} (\bibinfo{year}{1987}).

\bibitem[{\citenamefont{Hutson}(1994)}]{Hutson:cpc:1994}
\bibinfo{author}{\bibfnamefont{J.~M.} \bibnamefont{Hutson}},
  \bibinfo{journal}{Comput. Phys. Commun.} \textbf{\bibinfo{volume}{84}},
  \bibinfo{pages}{1} (\bibinfo{year}{1994}).

\bibitem[{\citenamefont{Hutson}(2007{\natexlab{b}})}]{Hutson:resfit:2007}
\bibinfo{author}{\bibfnamefont{J.~M.} \bibnamefont{Hutson}},
  \emph{\bibinfo{title}{Resfit 2007 computer program}}
  (\bibinfo{year}{2007}{\natexlab{b}}).

\bibitem[{\citenamefont{Johnson}(1973)}]{Johnson:1973}
\bibinfo{author}{\bibfnamefont{B.~R.} \bibnamefont{Johnson}},
  \bibinfo{journal}{J. Comput. Phys.} \textbf{\bibinfo{volume}{13}},
  \bibinfo{pages}{445} (\bibinfo{year}{1973}).

\end{thebibliography}
\end{document}